\documentclass[12pt,preprint,natbib,iop]{emulateapj}
\usepackage{graphicx,epsfig,hyperref,amsmath}

\lefthead{van der Wel et al.}
\righthead{Structural Parameters of Galaxies in CANDELS}
\begin{document}

\newcommand{\kms}{\>{\rm km}\,{\rm s}^{-1}}
\newcommand{\reff}{R_{\rm{eff}}}
\newcommand{\msol}{M_{\odot}}
\newcommand{\gf}{{\tt GALFIT}~}
\newcommand{\gala}{{\tt GALAPAGOS}~}
\newcommand{\mh}{H_{\rm{F160W}}}
\newcommand{\mj}{J_{\rm{F125W}}}
\newcommand{\my}{Y_{\rm{F105W}}}
\newcommand{\ser}{S\'ersic~}
\newcommand{\sext}{{\tt SExtractor}~}
\newcommand{\TT}{{\tt TinyTim}~}

\title{Structural Parameters of Galaxies in CANDELS}

\author{A.~van der Wel\altaffilmark{1},
E.F.~Bell\altaffilmark{2},
B.~H\"aussler\altaffilmark{3},
E.J.~McGrath\altaffilmark{4},
Yu-Yen Chang\altaffilmark{1},
Yicheng Guo\altaffilmark{5},
D.H.~McIntosh\altaffilmark{6},
H.-W.~Rix\altaffilmark{1},
M.~Barden\altaffilmark{7},
E.~Cheung\altaffilmark{4},
S.M.~Faber\altaffilmark{4},
H.C.~Ferguson\altaffilmark{8},
A.~Galametz\altaffilmark{9},
N.A.~Grogin\altaffilmark{8},
W.~Hartley\altaffilmark{3},
J.S.~Kartaltepe\altaffilmark{10},
D.D.~Kocevski\altaffilmark{4}, 
A.M.~Koekemoer\altaffilmark{8},
J.~Lotz\altaffilmark{8},
M.~Mozena\altaffilmark{4},
M.A.~Peth\altaffilmark{11},
Chien~Y.~Peng\altaffilmark{12}
}

\altaffiltext{1}{Max-Planck Institut f\"ur Astronomie, K\"onigstuhl
  17, D-69117, Heidelberg, Germany; e-mail:vdwel@mpia.de}

\altaffiltext{2}{Department of Astronomy, University of Michigan, 500
  Church Street, Ann Arbor, Michigan, 48109, USA}

\altaffiltext{3}{Schools of Physics \& Astronomy, University of
  Nottingham, University Park, Nottingham NG7 2RD, UK}

\altaffiltext{4}{UCO/Lick Observatory, Department of Astronomy and
  Astrophysics, University of California, Santa Cruz, CA 95064, USA}

\altaffiltext{5}{Astronomy Department, University of Massachusetts,
  710 N. Pleasant Street, Amherst, MA 01003, USA}

\altaffiltext{6}{Department of Physics \& Astronomy, University of
  Missouri-Kansas City, 5110 Rockhill Road, Kansas City, MO 64110,
  USA}

\altaffiltext{7}{Institute of Astro- and Particle Physics, University
  of Innsbruck, Technikerstraße 25, A-6020 Innsbruck, Austria}

\altaffiltext{8}{Space Telescope Science Institute, 3700 San Martin
  Drive, Baltimore, MD 21218, USA}

\altaffiltext{9}{INAF - Osservatorio di Roma, Via Frascati 33,
  I-00040, Monteporzio, Italy}

\altaffiltext{10}{National Optical Astronomy Observatory, 950
  N. Cherry Avenue, Tucson, AZ 85719, USA}

\altaffiltext{11}{Department of Physics and Astronomy, Johns Hopkins
  University, Baltimore, MD 21218, USA}

\altaffiltext{12}{Giant Magellan Telescope Organization 251, South
  Lake Avenue, Suite 300, Pasadena, CA 91101, USA}

\begin{abstract}
  We present global structural parameter measurements of 109,533
  unique, $\mh$-selected objects from the CANDELS multi-cycle treasury
  program.  \ser model fits for these objects are produced with \gf in
  all available near-infrared filters ($\mh$, $\mj$ and, for a subset,
  $\my$).  The parameters of the best-fitting \ser models (total
  magnitude, half-light radius, \ser index, axis ratio, and position
  angle) are made public, along with newly constructed point spread
  functions for each field and filter.  Random uncertainties in the
  measured parameters are estimated for each individual object based
  on a comparison between multiple, independent measurements of the
  same set of objects.  To quantify systematic uncertainties we create
  a mosaic with simulated galaxy images with a realistic distribution
  of input parameters and then process and analyze the mosaic in an
  identical manner as the real data. We find that accurate and precise
  measurements -- to 10\% or better -- of \emph{all} structural
  parameters can typically be obtained for galaxies with $\mh < 23$,
  with comparable fidelity for basic size and shape measurements for
  galaxies to $\mh\sim 24.5$.
\end{abstract}

\section{Introduction}
The Cosmic Assembly Near-Infrared Deep Extragalactic Legacy Survey
(CANDELS) constitutes 902 orbits of \textit{Hubble Space Telescope}
(HST) observing time.  One of the primary motivations for CANDELS is
the investigation of galaxies at redshifts $z\sim 1.5-3$, and in
particular their structural and morphological properties
\citep{grogin11}.  This paper describes the characterization of
structural galaxy properties in the HST WFC3/IR imaging mosaics
\citep{koekemoer11} through fitting the observed surface brightness
distributions by two-dimensional parametrized models, whose surface
brightness profiles follow the \ser law \citep{sersic68}.  The
resulting catalogs are made public online.

Over the previous decade WFPC2 and ACS have enabled comprehensive
structural and morphological studies of the galaxy population up to
$z\sim 1$ at `optical' wavelengths \citep[e.g.,][]{giavalisco04,
  rix04, scoville07b}.  At redshifts $z>1$ WFPC2 and ACS observations
sample the rest-frame ultra-violet, which complicates the
interpretation of galaxy images, as dust can strongly attenuate the
light and young stars that contribute little to the underlying stellar
mass distribution dominate over the older population.

For about a decade now, deep near-infrared surveys have been conducted
to remedy this \citep[e.g.,][]{labbe03, fontana04, quadri07,
  lawrence07, capak07}.  These ground-based surveys have provided
large samples, and despite the limited spatial resolution (typically
0.4-0.8\arcsec, or $\sim 3-6$ kpc at $z>1$), general and fundamental
structural properties have been measured out to $z\sim 2$
\citep{trujillo04, trujillo06, franx08, williams09, chang12}.
Galaxies were shown to be smaller in the past, and at all redshifts
their sizes were found to correlate with star-formation activity.  As
in the local universe, galaxies with low star-formation activity tend
to be smaller than galaxies with high (or normal) star-formation
rates. 

The use of ground-based near-infrared imaging for the purpose of
investigating the internal structure of galaxies and its evolution has
largely been limited to simple size measurements.  Any examination of
galaxy structure beyond this requires HST resolution.  Near-infrared
observations with NICMOS over relatively small areas and targeted
sampling of small numbers of pre-selected galaxies have before
revealed some fundamental aspects of structural properties at $z\sim
2$.  Beyond confirming the size evolution of galaxies, the profoundly
different structure of high-$z$ galaxies could now be fully
appreciated.  The discovery of massive, yet very small, galaxies
\citep{zirm07, toft07} posed a surprise to the community, and has
instigated much debate regarding the formation and evolution of the
most massive galaxies.  Moreover, the concentration of $z\sim 2$
galaxy light profiles, presumably tracing the bulge-to-disk ratio, was
found to correlate with star-formation activity, indicating that
galaxies with low star-formation rates have larger bulges at all
$z\lesssim 2$ \citep[e.g.,][]{kriek09}.

The arrival of WFC3 has now allowed the exploration of galaxy
structure at $z>1$ with unprecedented data quality and sample sizes.
The Ultra Deep Field (UDF) program \citep{bouwens10} broke ground in
terms of depth and resolution, confirming previous claims on the
structural properties of $z\sim 2$ galaxies and revealing further
detail on their morphology and structure \citep[e.g.,][]{szomoru10}.
Results from the HST/WFC3 Early Release Science (ERS) program
\citep{windhorst11} foreshadowed the power of CANDELS. That modestly
sized ERS program provided data quality and quantity on par with the
largest NICMOS data sets in existence \citep{scoville07b, conselice11}
and produced new insights into the basic structural parameters
\citep[e.g.,][]{vanderwel11, cassata11} and morphologies
\citep[e.g.,][]{cameron11} of $z\sim 2$ galaxies.

CANDELS fills the gap between the ground-based surveys and the
existing HST near-infrared surveys.  It provides much larger samples
than previous near-infrared HST data sets and much improved depth and
resolution compared to ground-based surveys.  \citet{wuyts11} and
\citet{bell12} showed with unprecedented clarity how star-formation
activity and structure are strongly related.  \citet{papovich12} and
\citet{lotz12} examined the environmental dependence of galaxy sizes
and merger frequency, respectively.  Moreover, the larger area allows
to study the properties of more rare objects.  \citet{kocevski12}
compared the morphologies of galaxies that do and do not host AGN and
concluded there is no significant difference, and \citet{kartaltepe12}
found a tentative connection between extreme star formation activity
and merging.

CANDELS also allows us to probe galaxy structure down to previously
unattainable low mass and luminosity limits -- apart from the UDF --
in particular through the deep segment of the survey.  Furthermore,
the structure of massive galaxies beyond $z=3$, where knowledge is
still sparse, can be explored \citep[e.g.,][]{caputi12, guo12}.
Finally, CANDELS aims at obtaining for the first time a comprehensive
view of the morphologies of $z>4$ Lyman-break and Ly-$\alpha$ emitters
at wavelengths longer than $2000\AA$ in the rest frame.

In this paper we describe the measurements of structural parameters of
109,533 unique objects in the CANDELS WFC3/IR data, representing
roughly 2/3 of the full survey. Our online materials will be updated
with the final 1/3 of the survey once observations have been completed
by the end of 2013.  We describe the imaging, object detection, and
background flux level estimation in \S\ref{sec:data}.  In
\S\ref{sec:psf} we construct Point Spread Function (PSF) models that
are used as the default PSFs in the CANDELS collaboration; these
models are also made public.  Using \gala \citep{barden12} we prepare
flux and noise image cutouts and describe how \gf \citep{peng10} is
used to produce single-component \ser model fits (\S\ref{sec:gf}).  We
estimate random and systematic uncertainties in the parameter
estimates through the comparison between different image data sets of
the same objects, and \ser fits to simulated galaxy images
(\S\ref{sec:err}).  We provide in \S\ref{sec:sum} a description of the
content of all published materials.

\begin{table}\scriptsize
  \caption{
    Publication dates of CANDELS catalogs with \gf-based structural parameters published online. Those in bold face are published along with this paper. The rest will be published over the next year, within 4 months after the last observation of each field.}\label{tab:list}
\begin{tabular}{cccc}
\hline
\hline
Field   & F105W       & F125W & F160W \\
\hline
COSMOS  & \nodata & \textbf{12/2012} & \textbf{12/2012} \\
EGS     & \nodata & 09/2013 & 09/2013 \\
GOODS-N & 12/2013 & 12/2013 & 12/2013 \\
GOODS-S & \textbf{12/2012} & \textbf{12/2012} & \textbf{12/2012} \\
UDS     & \nodata & \textbf{12/2012} & \textbf{12/2012} \\
\hline
\hline
\end{tabular}
\end{table}

\begin{figure*}[t]
\epsscale{.5} 
\plotone{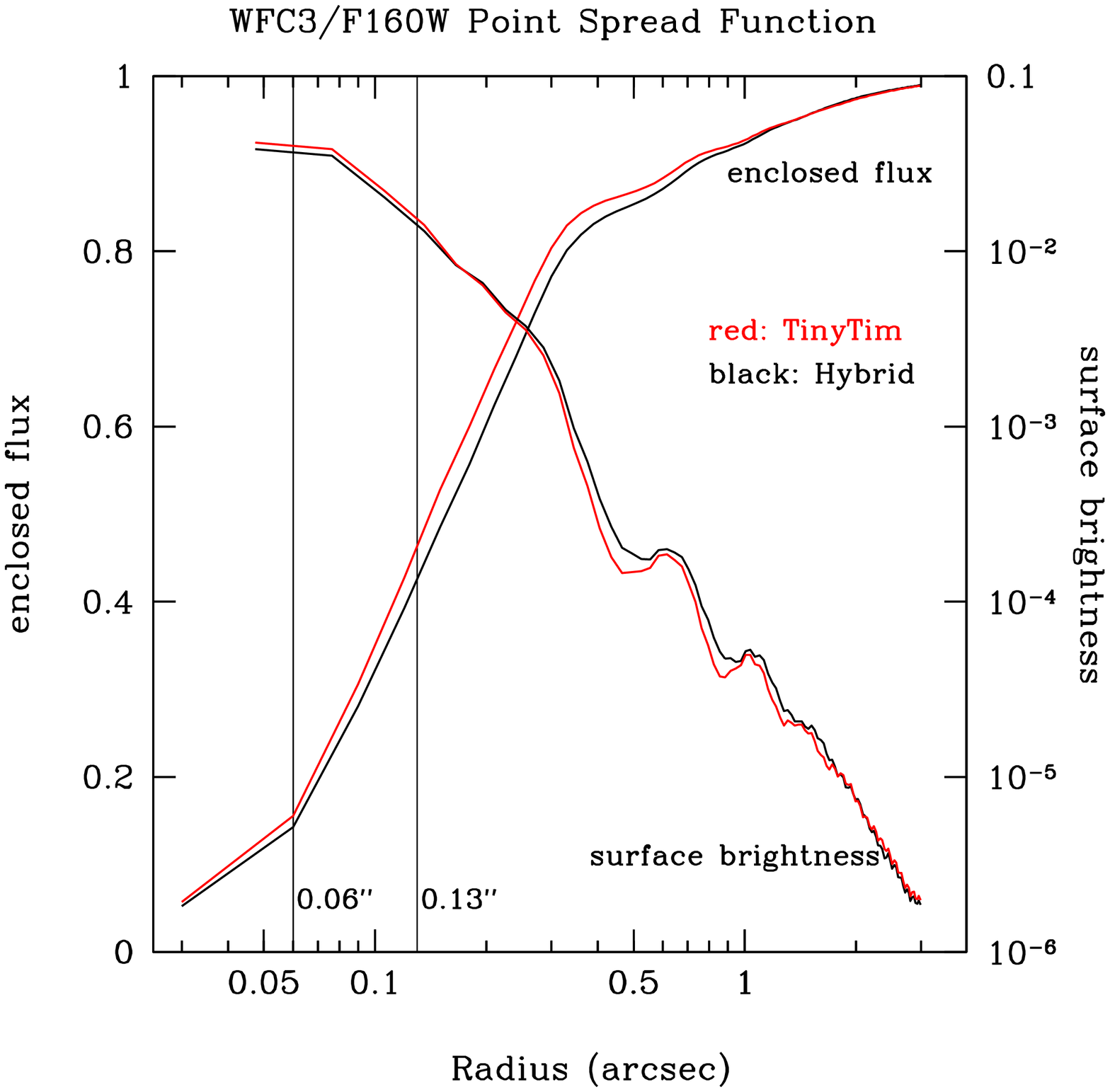}
\plotone{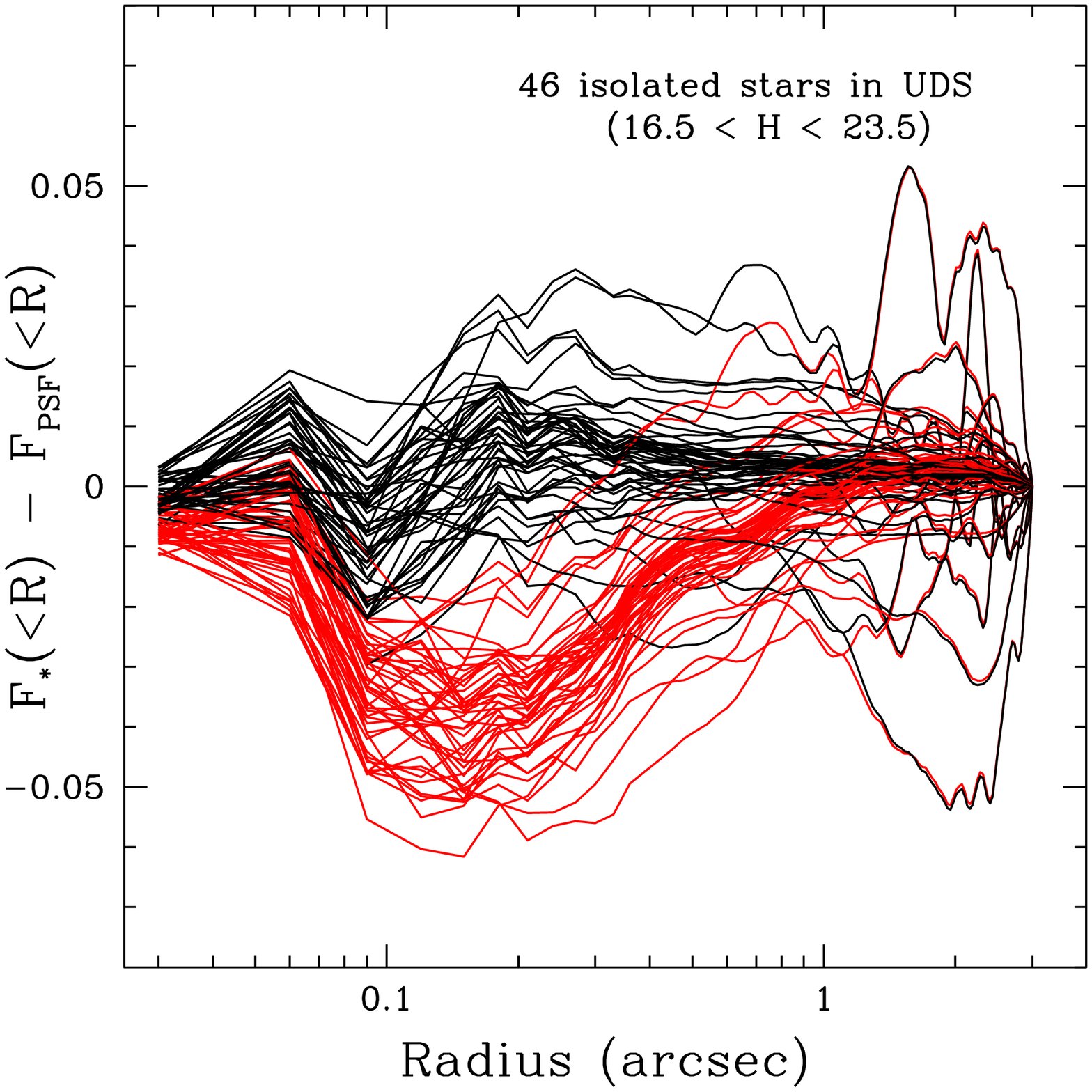}
\caption{ \textit{Left:} Comparison between two HST/WFC3 F160W model
  PSFs.  In red we show the \TT model PSF, in black a hybrid of a
  stacked star (outside a radius of 3 drizzled pixels, i.e.~,
  0.18\arcsec) and the \TT model (inside that radius).  The two
  sets of curves show the enclosed flux (left-hand y-axis) within the
  radius $r$ shown on the x-axis, and the surface brightness per pixel
  (right-hand axis) at radius $r$. The vertical lines indicate the
  original WFC3 pixel scale ($0.13\arcsec$) and the drizzled pixel
  scale of the mosaics used for the profile fitting in the paper
  ($0.06\arcsec$). The growth curves are normalized to the total flux
  of the model PSF within $10\arcsec$ and the growth curves of the
  empirical PSFs is forced to coincide with the that of the model PSF
  at $3\arcsec$.  The surface brightness curves are azimuthally
  averaged and in units of fractional PSF flux per $0.06\arcsec$
  pixel. \textit{Right:} The residuals from individual stars after
  subtracting the \TT model PSF (red) and the hybrid model PSF
  (black).  The systematic negative trend for the \TT model
  indicates that a larger fraction of the flux of a star resides at
  radii $0.2-1$\arcsec than what the \TT model predicts.  Our
  hybrid model PSF does not show such a trend and follows the light
  profiles of the isolated stars out to radii of 3\arcsec.}
\label{fig:psf}
\end{figure*}

\section{HST/WFC3 Imaging Mosaics}\label{sec:data}

\subsection{Fields and Filters}\label{sec:fields}

A full description of the CANDELS observing program is given by
\citet{grogin11} and \citet{koekemoer11}.  CANDELS is a WFC3 and
parallel ACS, 902 orbit HST imaging survey; here we concentrate on the
WFC3 data only, which covers 800 arcmin$^2$ and is distributed over
five widely separated fields.  There is a `deep' and a `wide'
component.  About 125 arcmin$^2$, divided over two fields, will have
$\sim$13 orbits per tile divided over three filters (F105W, F125W, and
F160W), and the remaining area, distributed over five fields, will
have $2-3$~orbits per tile divided over two filters (F125W and F160W).
These exposures reach $\mh\sim$28 and $\mh\sim$27 5$\sigma$ magnitude
limits for point sources in each filter for the `deep' and `wide'
imaging, respectively.

Along with this paper we electronically release structural parameter
catalogs for the three fields for which observations have been
completed at the time of writing, as summarized in Table
\ref{tab:list}, namely, the Ultra Deep Survey field
\citep[UDS,][]{lawrence07}, the Cosmological Evolution Survey field
\citep[COSMOS,][]{scoville07a} (both 9$\arcmin\times24\arcmin$ arcmin
and each at `wide' depth), and the Great Observatories Origins Deep
Survey-South field \citep[GOODS-S,][]{giavalisco04}: `wide' over
4$\arcmin\times10\arcmin$; `deep' over 7$\arcmin\times10\arcmin$.
Note that these dimensions refer to the CANDELS coverage, not to the
dimensions of the data sets that define the original surveys.

The CANDELS observations are augmented by previously obtained WFC3/IR
data from the Early Release Science program
\citep[ERS,][]{windhorst11} in the Northern part of the GOODS-S field
(4$\arcmin\times9\arcmin$ at 2-orbit depth in F098M, F125W, and F160W)
and the Ultra Deep Field (UDF) program \citep{bouwens10} embedded in
the GOODS-S deep area (1 pointing with $\sim$15 orbits in F105W and
F125W, and 28 orbits in F160W).  The electronically available catalogs
published along with this paper, derived from the finalized CANDELS
imaging products at the date of publication, will be updated upon
completion of the two remaining fields: GOODS-North and the Extended
Groth Strip \citep[EGS,][]{davis07}.  Image, weight (inverse
variance), and exposure time mosaics are prepared by drizzling the
individual exposures onto a grid with rescaled pixel sizes of 0.06$"$.
This procedure is described in detail by \citet{koekemoer11}.

\subsection{Object Identification}\label{sec:det}
We use a modified version of \sext \citep{bertin96} v2.5 to identify
objects in the CANDELS F160W mosaics.  We refer to Galametz et al.~(in
prep.) for a full description of the source extraction but the main
steps are as follows.  The first modification to \sext is that we
added a buffer between the isophotal area of each object and pixels
used to estimate the local background.  Note that this modification
affects object identification and segmentation map construction, but
the newly measured background is not used in the surface brightness
profile fits (see below).  The second modification is the removal of a
bug that previously allowed disconnected regions to have the same
value in the segmentation map, and therefore have the same
identification number.  Our modified version produces objects
consisting of adjacent pixels only.

\sext is run in dual mode, using the F160W mosaics both as the
detection images and the measurement images.  Indeed we found slight
differences with respect to single-mode \sext output, which we wish to
avoid in order to allow direct comparisons with dual-mode photometry
of objects detected in F160W and measured in other filters.

Catalogs for two sets of \sext detection parameters, the `cold' setup
and the `hot' setup are produced separately and then combined in an
approach introduced by \citet{barden12}.  The `cold' setup focuses on
optimal segmentation of relatively bright objects.  This avoids that
star-forming galaxies are spuriously split up into multiple components
by deblending too aggressively.  The `cold' setup selects objects with
0.75$\sigma$ detections over 5 adjacent pixels after smoothing with a
top-hat kernel with a diameter of 9 pixels.  Objects are deblended
adopting 9 logarithmic sub-thresholds relative to the maximum count
value of an object and using a minimum contrast of 0.0001.  Given that
the PSF has a FWHM much smaller than that of the smoothing kernel used
in the `cold' setup, this approach is suboptimal for detecting small,
faint objects.

This is mitigated by the `hot' setup, which uses a Gaussian smoothing
kernel with a FWHM of 4 pixels that is similar to that of the PSF and
selects objects with 10 adjacent pixels with 0.7$\sigma$ fluxes.
Deblending is done with the default \sext parameters: a minimum
contrast of 0.001 and 64 logarithmic sub-thresholds.

Each object in the `hot' catalog that falls within an aperture around
an object in the `cold' catalog is considered part of the latter and
removed from the combined catalog.  This elliptical aperture is 2.5
times the elliptical Kron radius as measured with the default \sext
parameters.  The size and shape of the ellipse are calculated from the
second-order moment of the flux distribution.  In the combined
segmentation map the segmented pixels of the removed `hot' object are
added to the object in the 'cold' segmentation map.

In total, we identify 109,533 unique objects in the F160W mosaics.
HST resolution enables us to resolve close companions, which is
essential when one is interested in the measurement of the structures
of individual galaxies.  7\% (9\%) of objects in CANDELS `wide'
(`deep') have one or more neighbors within 1$\arcsec$; they would
typically not be resolved in ground-based imaging data sets.

The \sext measurements of object magnitudes, size, shape and
orientation are used to provide \gf with initial guesses for the
fitting parameters (see \S\ref{sec:gfsetup}).  This helps in terms of
computing time and ensures that \gf converges to the global $\chi^2$
minimum in parameter space.

\begin{figure*}[t]
\epsscale{1.1} 
\plotone{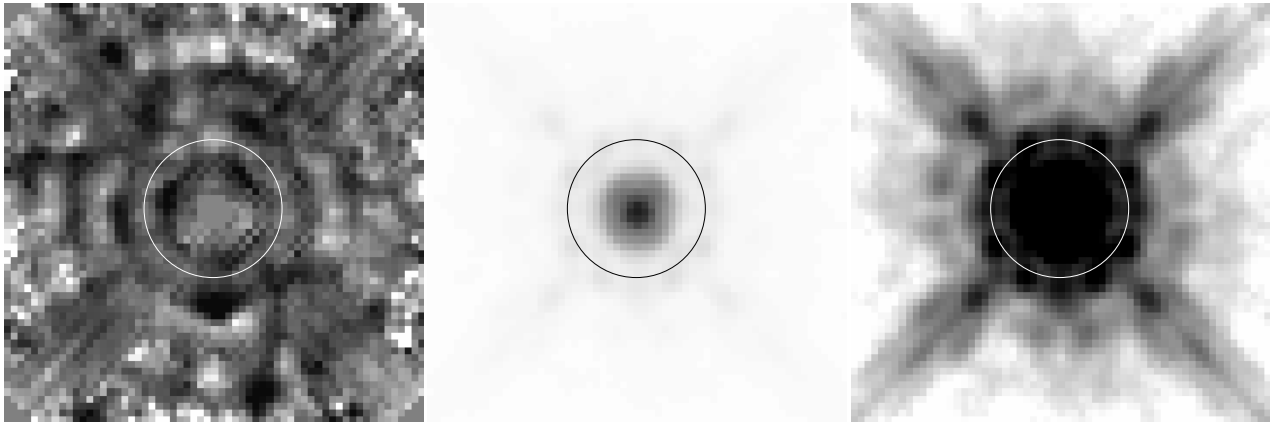}
\caption{ \textit{Left:} Ratio of our hybrid PSF and the \TT
  model, both for the UDS field, with dark regions lower levels in the
  \TT model (the darkest regions have values $>2$).  The image is
  3.6\arcsec$\times $3.6\arcsec, and the circle, which has been
  included to guide the eye, has a radius of 0.6\arcsec. The area near
  the center has a constant value by construction as we use the
  \TT model in our hybrid for the central pixels.  We show our
  hybrid PSF with two different stretches in order to to emphasize the
  bright, central region (\textit{Middle}) and the faint, outer region
  (\textit{Right}).  At each radius, the brightest parts of the PSF
  are reproduced quite accurately by the \TT model; rather, it is
  in between the diffraction spikes and Airy rings that the deviations
  are largest, and at a level that cannot be explained by noise. These
  features cause the systematic differences between the light profiles
  of stars and the \TT model shown in Figure \ref{fig:psf}.}
\label{fig:psf2d}
\end{figure*}

\subsection{Background Estimation}\label{sec:bg}
The image cutouts we use for model fitting (\S\ref{sec:prep}) are too
small for \gf to optimally determine the background flux level.  In
order for the image to be large enough for an accurate, simmulataneous
determination of both the structural parameters and the background
flux level, the number of objects that would have to be fit
simultaneously would become impractical, if not impossible in some
cases (at present, \gf cannot fit more than 110 objects at one time).

The pipeline
\gala\footnote{http://astro-staff.uibk.ac.at/$\sim$m.barden/galapagos/}
\citep{barden12} remedies this problem.  Originally written to analyze
ACS mosaics from GEMS \citep{rix04}, \gala determines the background
from the full mosaic and then runs \gf using this background value
along with initial guesses for the input parameters based on the \sext
output described above.

The independent background estimate is the main feature of \gala that
elevates it beyond the level of `just' a smart wrapper for \sext and
\gf.  In short, ignoring pixels within a certain distance of an
object, the flux is computed in a series of annuli, searching for a
converging flux level at sufficiently large distances from the target
object.  This background flux level is provided to \gf, which keeps it
fixed at that value.  In \S\ref{sec:random} we investigate the
contribution of the uncertainty in the background to the uncertainties
in the structural parameters.

\section{Point Spread Function}\label{sec:psf}
Besides an accurate noise model and background estimation, good
knowledge of the PSF is essential for the accuracy of the parameter
estimation.  This necessitates the production of custom-made PSF
models for each of the mosaics and each of the filters.  These PSF
models should be used for all PSF-dependent analyses of the public
CANDELS imaging mosaics, and are therefore released here (see
\S\ref{sec:sum} for a summary of the published material).

The PSF has a FWHM of $\sim$0.17\arcsec~for the F160W filter.
Individual exposures (typically 800s) are dithered to improve on the
originally coarse sampling -- the pixel scale of the WFC3/IR camera is
0.13\arcsec.  Using the \TT package \citep{krist95} we construct a
model PSF for the respective imaging mosaics for the different fields
and filters.  Briefly, \TT PSF models are created for the center of
the WFC3 detector assuming a G2V spectral type.  The PSFs are
sub-sampled by a factor of ten in order to aid in aligning them with
the CANDELS dither pattern.  They are then re-sampled to the WFC3
pixel scale and a kernel is applied to replicate the effects of
inter-pixel capacitance \citep{hilbert11}.  Finally, PSFs at each
dither position are distortion-corrected and combined with the same
drizzle parameters used in producing the imaging mosaics
\citep{koekemoer11}.

As an example we show the growth curve and surface brightness profile
of the F160W PSF for the UDS (Figure \ref{fig:psf} -- red lines in the
left-hand panel). To verify the accuracy of this model PSF we subtract
it from a sample of 46 isolated stars in the F160W UDS mosaic (after
normalizing the star fluxes within a radius of 3\arcsec, and examine
the residual enclosed flux (Figure \ref{fig:psf} -- red lines in the
right-hand panel).  The systematic residual of $\sim-4$\% seen at
$\sim$0.2\arcsec~implies that the model PSF contains $\sim$4\% less
light outside an aperture $\sim0.2$\arcsec~than point sources in the
actual data. Background levels, computed as described in
\S\ref{sec:bg}, are low compared to the stellar fluxes, even at radii
of $1-3$\arcsec, such that the enclosed flux is affected by less than
1\% at any radius.  In other words, background errors cannot explain
the large differences between stellar and model PSF growth curves at
relatively small radii.  Such an erroneous PSF model, which is seen
for other fields and filters as well, can affect the measurement of
structural parameters of small, concentrated objects at a level that
exceeds the formal, random uncertainty.

The difference between the stars and the \TT model occurs in
relatively `dark' areas seen in Figure \ref{fig:psf2d}.  The center,
the diffraction spikes, and the dot-like features associated with the
first Airy ring are well reproduced by \TT, whereas regions in between
those features are too dark. The origin of this feature remains
unexplained.

We use the 46 isolated stars shown in the right-hand panel of Figure
\ref{fig:psf} in median-stacked form as an alternative PSF
representation.  The problem with this approach is the variety in the
sub-pixel positioning of the stellar images, which leads to broadening
in the PSF model when stacking a number of stars.

In order to provide an accurate PSF model at all radii we take the
median-stacked star and replace the central pixels (within a radius of
3 pixels from the center) by the \TT model PSF.  The flux values of
these pixels are normalized such that the total flux of the newly
constructed hybrid PSF model is the same as that of the stacked star.
The accuracy of the model PSF in the central region is confirmed by
its good correspondence with the flux distributions of the small
number stars that happen to be well-aligned with the brightest pixel.
Note that there are too few such stars to produce a stacked PSF with
sufficient signal-to-noise ratio at larger radii.

The growth curves, surface brightness profiles and residuals are shown
in black in Figure \ref{fig:psf}.  The residuals show much smaller
systematic effects than the best-effort \TT model PSF, indicating that
the hybrid PSF provides a good description of the growth curves of
individual stars out to 3\arcsec.

We construct equivalent hybrid PSF models for all fields and filters.
For the UDS and COSMOS fields we created PSF models for the F125W and
F160W PSF filters.  For GOODS-S the situation is more complex.  For
the ERS region we have F098M, F125W and F160W PSF models; for the
`wide' and `deep' regions we created separate F105W, F125W and F160W
PSF models; for the UDF, which is very small and contains few stars,
we simply make use of a single star.  Replacing the central pixels by
model values does not work well for a single star, but the chosen star
is bright and isolated, and happens to be be relatively well-centered
on the brightest pixel, and thus serves its purpose well.

The fact that variations are $\lesssim 2$\% for all stars implies that
PSF differences due to variations in position, color, and magnitude
are negligible for our \gf measurements.  We note that small
variations in dither pattern, orientation and general mosaic geometry
exist within the distinct fields.  This implies that some fraction of
the objects will not have precisely the correct PSF model.  This is
unavoidable when using stars to produce the PSF model.

\begin{figure*}[t]
\epsscale{1.2} 
\plotone{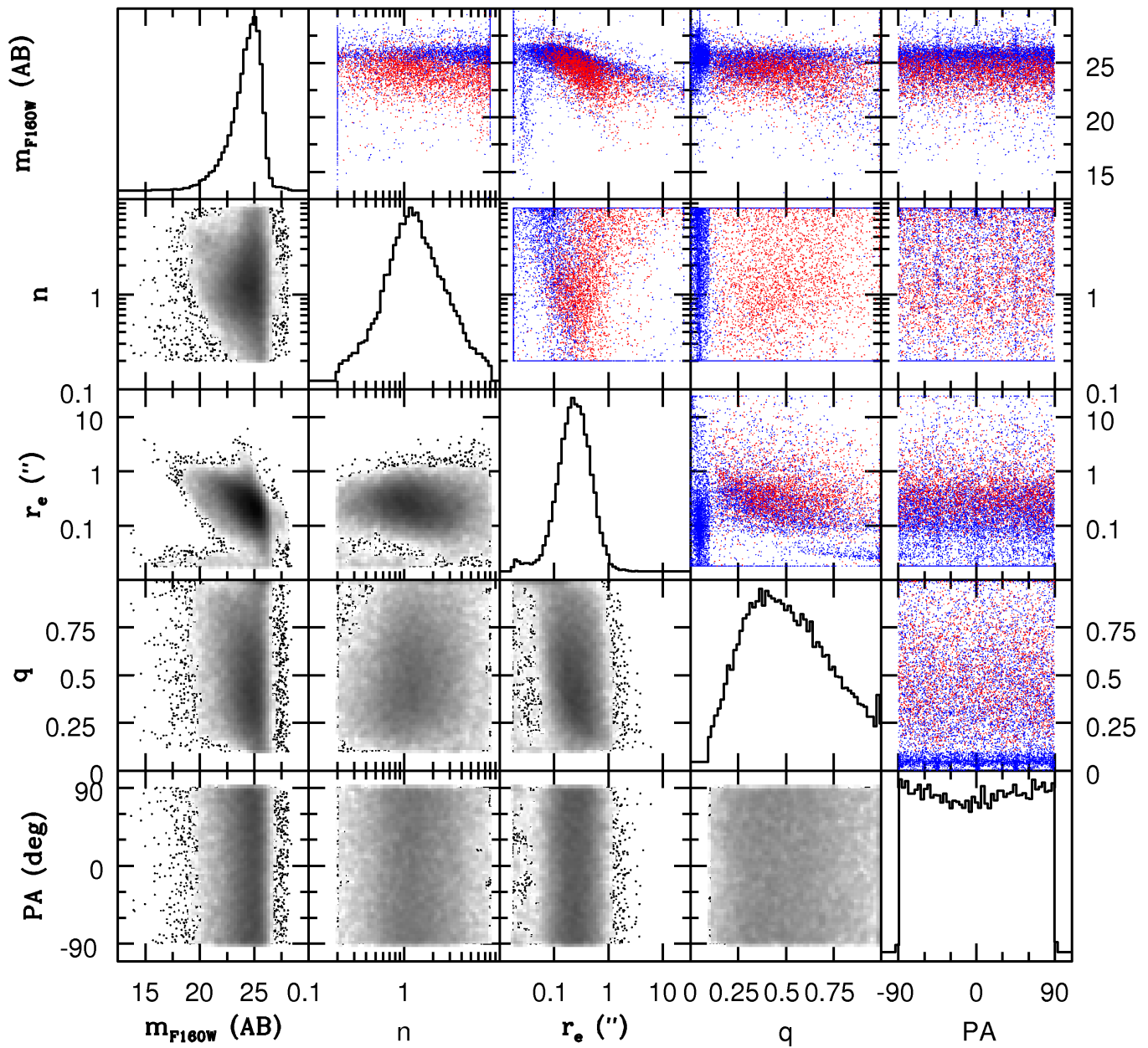}
\caption{ The distribution of the \gf-derived parameters F160W
  magnitude, \ser index, effective radius, axis ratio, and position
  angle for objects in the three CANDELS fields analyzed here (COSMOS,
  GOODS-South, and UDS).  The panels with gray scale distributions
  represent the objects with flag value zero (these are objects with
  good fits -- see \S\ref{sec:flags}). On the diagonal we show for the
  flag zero sample the histogram of each of the five parameters in
  units as shown on the x-axis.  The typical galaxy has magnitude
  $\mh\sim 25$, $n\sim 1$, $r_e\sim 0.3\arcsec$, and $q\sim 0.4$. The
  colored panels show the distribution of objects with flag value one
  (red) and flag value two (blue) -- these are objects with,
  respectively, suspicious and bad fits as explained in
  \S\ref{sec:flags}.}
\label{fig:summary}
\end{figure*}

\begin{figure}[t]
\epsscale{1.2} 
\plotone{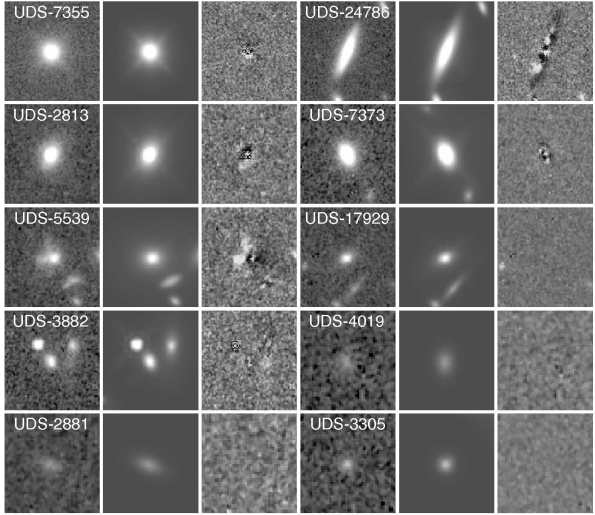}
\caption{ Ten galaxy images (\emph{left}) from the UDS F160W mosaic
  and their \gf models (\emph{middle}) and residuals (\emph{right}).
  These objects have been chosen to have a range in magnitude
  (increasing from $\mh\sim 20$ at the top left to $\mh \sim 25$ at
  the bottom right) and structure. UDS-7335 and UDS-2813 have $n\sim
  4$ and smooth light profiles characteristic of early-type galaxies;
  UDS-24786 and UDS-7373 have $n\sim 1$ and show more substructure,
  characteristic of star-forming, late-type galaxies.  For the
  faintest galaxies shown here the \gf model still captures the basic
  structural parameters such as size and shape.  In several cases
  multiple objects are fit simultaneously, illustrating the procedure
  described in \S\ref{sec:gfsetup}.}
\label{fig:stamps}
\end{figure}

\begin{figure}[t]
\epsscale{1.2} 
\plotone{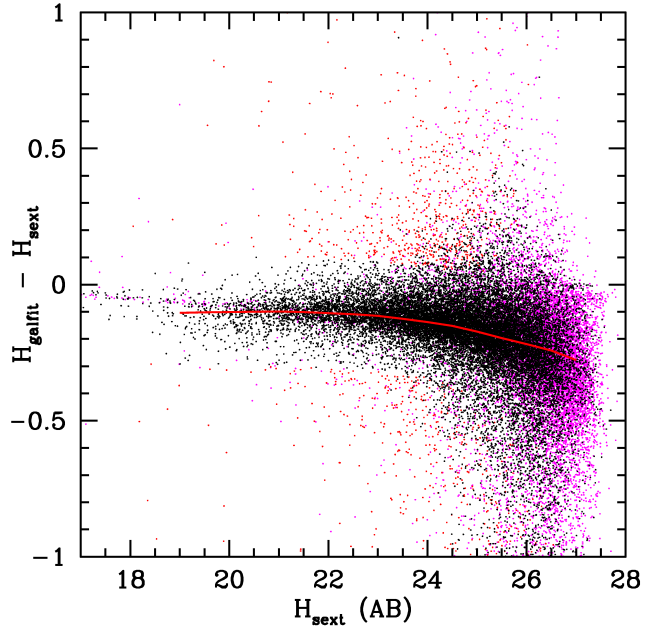}
\caption{ Comparison of \gf and \sext (BEST) F160W magnitudes for
  objects in the UDS.  The y-axis shows the difference between the
  two.  Black points represent objects that we have assigned flag
  value 0 (good fits); red points have flag value 1 (suspect fits);
  magenta points have flag value 2 (bad fits -- bright stars have bad
  fits because they are point sources).  See \S\ref{sec:flags} for a
  full explanation of the flag definitions.  The red line is a running
  median for the black and red points, quantifying the systematic
  offset between \gf and \sext total magnitudes.  Note that these
  systematic offsets are in excess of the random uncertainty in the
  \gf-based total magnitudes.}
\label{fig:magmag}
\end{figure}

\begin{figure}[t]
\epsscale{1.2} 
\plotone{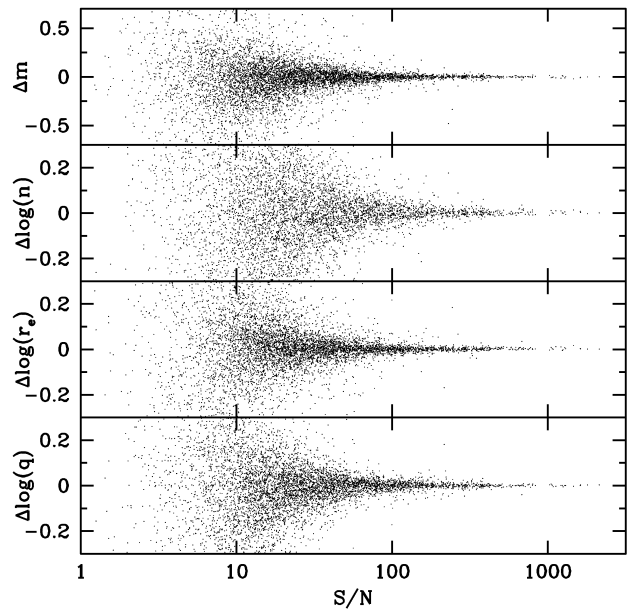}
\caption{ Difference between the structural parameter estimates from
  deep and shallow images of 6492 objects in GOODS-S as a function of
  $S/N$ in the shallow up.  The spread reflects the uncertainty in the
  measurements, which we use in \S\ref{sec:random} to assign
  uncertainties to the structural parameter measurements of all
  galaxies.  Although dependencies on other parameters exist (see
  Figures \ref{fig:correrr} and \ref{fig:correrr2}), the uncertainty
  depends to first order on $S/N$.}
\label{fig:sn_par}
\end{figure}

\begin{figure*}[t]
\epsscale{1.2} 
\plotone{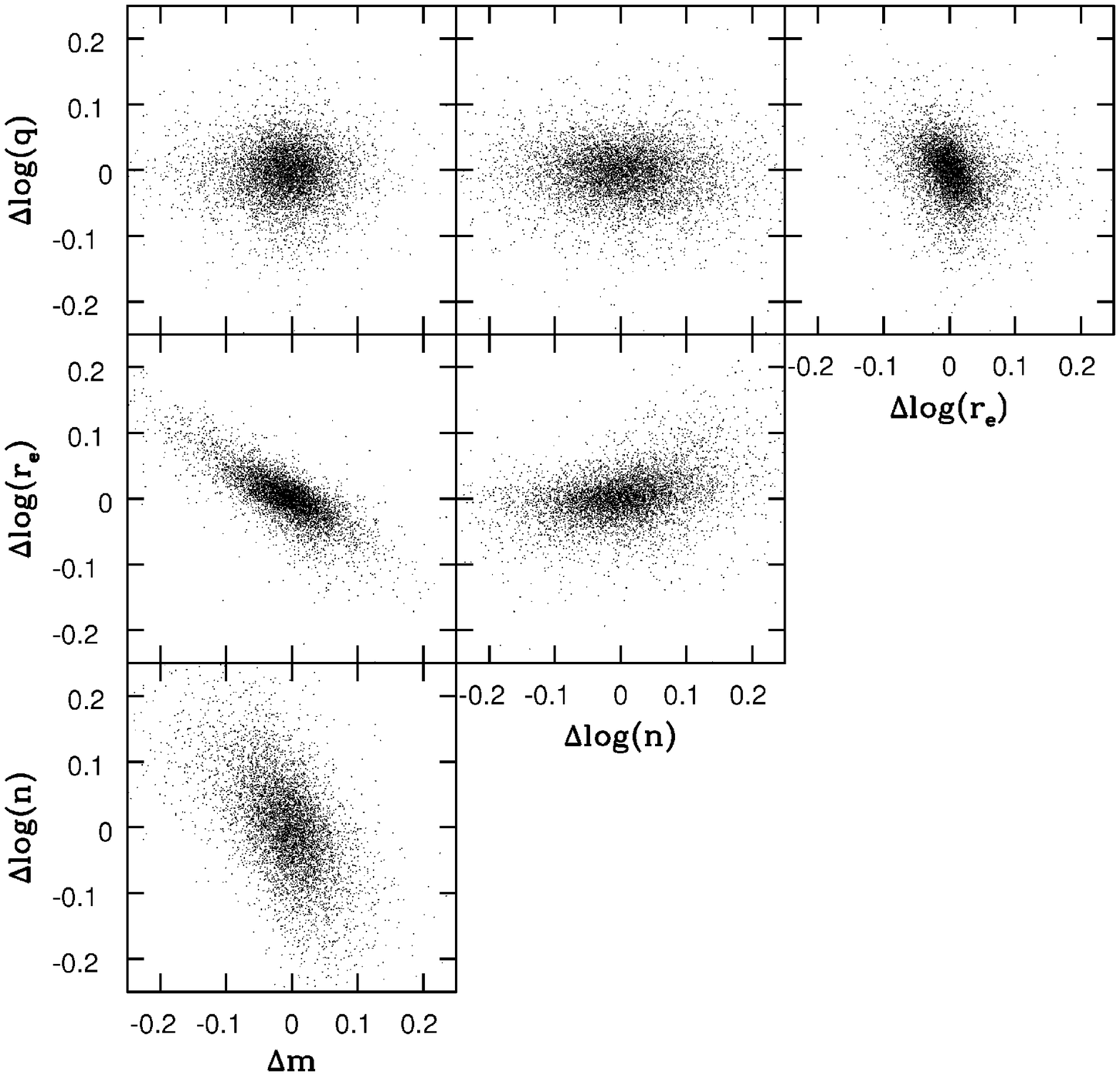}
\caption{ Correlations among structural parameter measurement
  uncertainties \citep[also see, e.g.,][]{haussler07, guo09,
    bruce12}. All quantities $\Delta$ are the difference between the
  structural parameters are measured from the deep and shallow images
  of 6492 galaxies.  $\Delta$ is normalized to $S/N=50$ as described
  in the text; that is, first-order effects on $S/N$ are removed.
  Particularly relevant are the strong correlations between the
  uncertainties in magnitude ($m$), \ser index ($n)$ and effective
  radius ($r$): uncertainties in $m$ and $r$ correlate such that any
  overestimate (underestimate) in $r$ corresponds linearly with an
  overestimate (underestimate) in the total flux; analogously, if the
  total flux is overestimated (underestimated) by 5\%, the \ser index
  is typically overestimated (underestimated) by as much as 25\%.  The
  shape and amplitude of these `error ellipses' (or rather,
  `ellipsoids', as they are at least 3-dimensional) depend on the
  parameters themselves and therefore do not represent the covariance
  matrix of the uncertainties for the sample as a whole.}
\label{fig:correrr}
\end{figure*}

\section{Modeling Light Profiles}\label{sec:gf}

\subsection{Preparation of Image and Noise Cutouts}\label{sec:prep}
For each of the 109,533 objects, an image cutout, in units of
electrons per second, is produced from the large mosaic. The size of
each rectangular cutout is determined by the Kron radius, ellipticity
and position angle as measured by \sext, and is chosen to enclose an
ellipse with major axis length 2.5 times the Kron ellipse ($A_{IMAGE}
\times KRON_{RADIUS}$ in \sext parlance).  This typically corresponds
to $\sim 30$ times the half-light radius of an object.

A noise map of the same size is also produced.  The pixel values in
the noise map have units of electrons per second, and are the square
root of the total variance. The total variance is estimated by
starting with the computed variance, referred to by
\citet{koekemoer11} as the `intrinsic' term (e.g., background noise
and readout noise), in units of electrons.  We add the variance at
each pixel from the Poisson noise due to the objects themselves.  We
divide by the (computed) exposure time for each pixel to make the
units consistent with those of the images.

The images and noise maps are used by \gf as inpu.  We note that \gf
does not have all necessary information about the image
characteristics required for a self-contained noise-estimate as a
result of the extensive process from raw, single exposures, to final,
drizzled mosaics.  Therefore, especially for bright, compact objects
it is essential to use the noise maps we construct rather than the
\gf's own noise estimation.  In practice, for objects brighter than
$\mh \sim 22-23$ -- corresponding to the typical background WFC3/F160W
flux level measured over an area the size of a typical object -- the
extrinsic noise term begins to dominate.

We compute signal-to-noise ratio ($S/N$) for each object from their
images and noise maps, integrating over the pixels that belong to the
objects according to the segmentation map.  An object with $\mh \sim
22$ has $S/N \sim 100-200$ for the `wide' data.  These $S/N$ estimates
are used in our derivation of the measurement uncertainties in
\S\ref{sec:err}.

\begin{figure*}[t]
\epsscale{1} 
\plotone{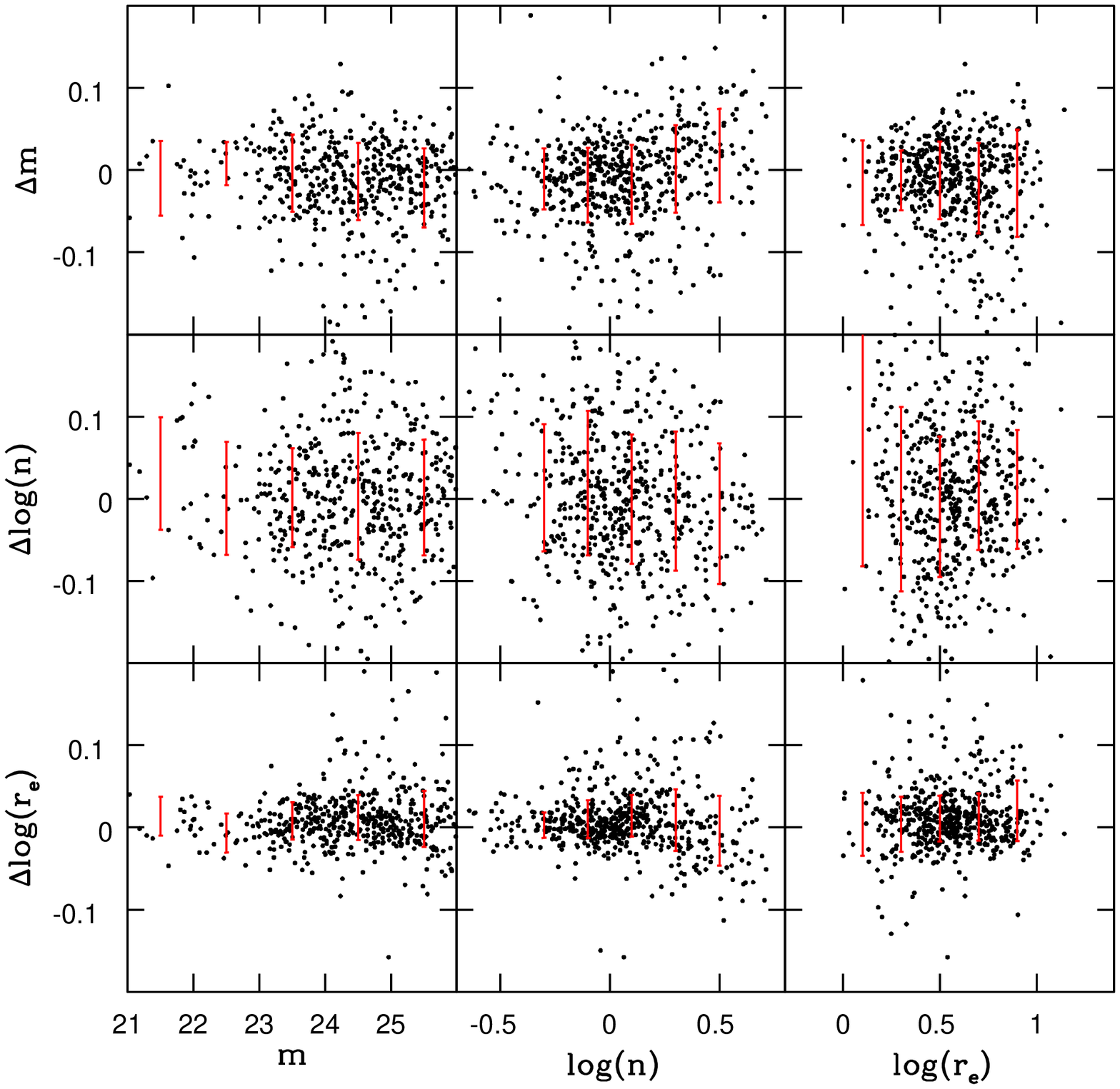}
\caption{ Correlation between structural parameter measurement
  uncertainties and measured structural parameters.  On the x-axis
  three parameters (magnitude, $m$, \ser index, $n$, and effective
  radius, $r$) as measured from the GSD deep region are shown.  The
  galaxies displayed in the first column of panels (showing $m$), are
  chosen within narrow ranges of $n$ and $r$ around the median values
  in the full catalog. In the second (third) column the same is done
  for $n$ ($r$): $m$ and $r$ ($n$) are chosen within a narrow range.
  On the y-axis the difference $\Delta$ between the measurements from
  the deep and shallow images of the same objects are shown: the
  scatter in $\Delta$ reflects the uncertainty, indicated in red by
  the running 16-84\%-tile ranges.  As in Figure \ref{fig:correrr},
  all $\Delta$ are normalized to $S/N=50$. Strong second-order effects
  remain, most clearly seen in the dependence of the uncertainty in
  $r$ and $m$ on $n$ and the dependence of the uncertainty in $n$ and
  $m$ on $r$.}
\label{fig:correrr2}
\end{figure*}

\subsection{\gf  Setup}\label{sec:gfsetup}
The image cutouts, along with the noise model, the appropriate PSF
model, and the pre-determined, fixed background level, are provided to
\gf, which is used to find the best-fitting \ser model for each
object.  The fitting parameters are total magnitude ($m$), half-light
radius ($r$) measured along the major axis, \ser index ($n$), axis
ratio ($q$), position angle ($PA$), and central position.  Initial
guesses for these parameters are taken from the \sext detection
catalog.  A constraints file is constructed so that \gf is forced to
keep the \ser index between 0.2 and 8, the effective radius between
0.3 and 400 pixels, the axis ratio between 0.0001 and 1, the magnitude
between 0 and 40, and, between -3 and +3 magnitudes from the input
value (the \sext magnitude).

Neighboring objects in each image cutout are fit simultaneously or
masked out, depending on their brightness compared to the main target:
galaxies are fit simultaneously if they are less than 4 magnitudes
fainter than the main target; stars are fit simultaneously if they are
less than 2 magnitudes fainter.  In order to produce a good model, it
is sometimes necessary to also fit objects outside the image cutout.
The decision to do so depends on the contribution of objects outside
the image cutout to the flux in the cutout.  \sext or previously
obtained \gf measurements are used to make informed decisions.  The
entire decision process, carried out by \gala, is quite sophisticated
and we refer to \citet{barden12} for a full description.  Note that
the segmentation maps described in \S\ref{sec:det} are only used to
identify objects, and not to identify the pixels that are used in the
\ser~fits.

For illustrative purposes, the images of 10 galaxies and their fits
are shown in Figure \ref{fig:stamps}.  All best-fitting \gf parameters
are listed in Table \ref{tab:data} and shown in Figure
\ref{fig:summary}.  The typical galaxy with a good \ser model fit (see
\S\ref{sec:flags} for the definition of `good') has magnitude $\mh\sim
25$, \ser index $n\sim 1$, effective radius $r_e\sim 0.3\arcsec$, and
axis ratio $q\sim 0.4$.  The position angle distribution is not
entirely uniform; this is due to unresolved sources, mostly stars.
For resolved sources the distribution is uniform.
   
\begin{table*}\scriptsize
  \caption{ 
    \gf fitting results and derived uncertainties.  
    This table gives a sample of the UDS F160W catalog.  
    Electronic versions of the full catalogs for all fields and (near-infrared) filters, as listed in Table \ref{tab:list}, are published online (see \S\ref{sec:sum}.
    $m$ is the AB magnitude, $r$ is the half-light semi-major axis in arcseconds, 
    $n$ is the \ser index, 
    $q$ is the projected axis ratio, $PA$ is the position angle in degrees ($PA=0$ corresponds to North; $PA=90$ corresponds to East).
    $S/N$ is the signal-to-noise ratio measured over the pixels in the segmentation map that are attributed to the object.
    The listed uncertainties are based on population statistics and derived as described in \S\ref{sec:random} and 
    correspond to the half-width of the 68\%-confidence interval (i.e., these are 1$\sigma$ error bars in the case of gaussianity).
  }
\begin{tabular}{cccccccccc}
  \hline
  \hline
  ID & RA & DEC & FLAG & $m \pm \delta m$ & $r \pm \delta r$ & $n \pm \delta n$ & $q \pm \delta q$ & $PA \pm \delta PA$ & $S/N$ \\ 
  \hline
  & J2000 & J2000 & & AB mag & $\arcsec$ & & & $\deg$ & \\
  \hline
  1 & 34.223766 & -5.278053 & 3 & -999  $\pm$ -999 &  -999 $\pm$ -999 &  -999 $\pm$ -999 &  -999 $\pm$ -999 & -999  $\pm$ -999   &  5.43 \\
  2 & 34.223904 & -5.277949 & 0 & 24.41 $\pm$ 0.40 &  0.43 $\pm$ 0.20 & 0.80  $\pm$ 1.03 & 0.18  $\pm$ 0.09 &  41.61 $\pm$ 14.32 &  6.88 \\
  3 & 34.223492 & -5.277952 & 0 & 25.17 $\pm$ 0.24 &  0.18 $\pm$ 0.06 & 0.27  $\pm$ 0.28 & 0.74 $\pm$ 0.37  & -52.04 $\pm$ 23.24 &  8.57 \\
  4 & 34.265106 & -5.277749 & 1 & 23.06 $\pm$ 0.19 &  0.27 $\pm$ 0.07 & 1.49  $\pm$ 0.84 & 0.72 $\pm$ 0.19  &  67.92 $\pm$ 10.47 & 13.33 \\
  5 & 34.295372 & -5.277648 & 1 & 21.06 $\pm$ 0.09 &  1.16 $\pm$ 0.11 & 1.55  $\pm$ 0.28 & 0.80 $\pm$ 0.08  & -62.85 $\pm$ 3.57  & 33.46 \\
  . & . & . & . & . & . & . & . & . & .\\
  . & . & . & . & . & . & . & . & . & .\\
  . & . & . & . & . & . & . & . & . & . \\
  \hline
  \hline
\end{tabular}
\label{tab:data}
\end{table*}

\begin{table*}\scriptsize
  \caption{
    Systematic and random uncertainties in structural parameters for CANDELS `wide' imaging in F160W. The
    systematic uncertainties are derived from \gala processing of
    simulated images as described (see \S\ref{sec:sys}); the random
    uncertainties are inferred from \gala processing of different data
    sets of the same galaxies as described in \S\ref{sec:random} .
    The systematic uncertainties are given first, followed by random
    uncertainties as error bars.  For $m$ the units are in magnitudes; for
    the other parameters the uncertainties are relative, in linear units
    (they correspond to percentages).  
    The galaxy samples are split by \ser 
    index (at $n=3$) and effective radius (at $r_e=0.3"$).
  }

\begin{tabular}{c|cccc|cccc}
\hline
\hline 
& \multicolumn{4}{c}{$\Delta m$} & \multicolumn{4}{c}{$\Delta n$} \\
 $m$ &   $n<3$ & $n>3$ & $r_e<0.3"$ & $r_e>0.3"$ &   $n<3$ & $n>3$ & $r_e<0.3"$ & $r_e>0.3"$  \\
\hline
21 & $-0.02\pm0.01$ & $-0.01\pm0.01$ & $-0.02\pm0.01$ & $-0.01\pm0.01$ & $-0.01\pm0.02$ & $-0.03\pm0.03$ & $-0.02\pm0.02$ & $-0.00\pm0.02$ \\
22 & $-0.01\pm0.02$ & $0.00\pm0.02$ & $-0.01\pm0.01$ & $-0.01\pm0.02$ & $-0.03\pm0.04$ & $-0.06\pm0.06$ & $-0.05\pm0.04$ & $-0.01\pm0.05$ \\
23 & $-0.01\pm0.03$ & $0.06\pm0.07$ & $-0.01\pm0.03$ & $0.01\pm0.05$ & $-0.01\pm0.10$ & $-0.09\pm0.16$ & $-0.01\pm0.09$ & $-0.03\pm0.13$ \\
24 & $-0.01\pm0.06$ & $0.13\pm0.13$ & $-0.01\pm0.05$ & $0.05\pm0.12$ & $-0.01\pm0.24$ & $-0.25\pm0.33$ & $-0.01\pm0.23$ & $-0.11\pm0.33$ \\
25 & $-0.04\pm0.13$ & $0.10\pm0.21$ & $-0.03\pm0.12$ & $0.03\pm0.25$ & $0.02\pm0.49$ & $-0.49\pm0.58$ & $0.00\pm0.45$ & $-0.20\pm0.68$ \\
26 & $-0.14\pm0.26$ & $0.02\pm0.31$ & $-0.13\pm0.25$ & $-0.09\pm0.44$ & $0.31\pm0.86$ & $-0.84\pm0.92$ & $0.23\pm0.84$ & $-0.16\pm1.21$ \\
27 & $-0.47\pm0.41$ & $-0.37\pm0.54$ & $-0.46\pm0.42$ & \nodata & $0.99\pm1.28$ & $-1.59\pm1.46$ & $0.87\pm1.29$ & \nodata \\
\hline
\hline
& \multicolumn{4}{c}{$\Delta r_e$} & \multicolumn{4}{c}{$\Delta q$} \\
\hline
 $m$ &   $n<3$ & $n>3$ & $r_e<0.3"$ & $r_e>0.3"$ &   $n<3$ & $n>3$ & $r_e<0.3"$ & $r_e>0.3"$  \\
\hline
21 & $0.00\pm0.01$ & $0.00\pm0.016$ & $0.01\pm0.01$ & $0.00\pm0.01$  & $0.01\pm0.01$  & $-0.01\pm0.02$ & $0.01\pm0.01$  & $0.0\pm0.01$   \\
22 & $0.01\pm0.02$ & $-0.02\pm0.04$ & $0.02\pm0.02$ & $-0.01\pm0.02$ & $0.00\pm0.02$ & $0.01\pm0.03$  & $0.00\pm0.02$  & $0.0\pm0.03$   \\
23 & $0.00\pm0.04$ & $-0.10\pm0.11$ & $0.00\pm0.03$ & $-0.03\pm0.06$ & $0.00\pm0.05$  & $-0.01\pm0.08$ & $0.00\pm0.04$  & $0.00\pm0.06$  \\
24 & $0.01\pm0.08$ & $-0.22\pm0.19$ & $0.01\pm0.08$ & $-0.10\pm0.15$ & $-0.02\pm0.11$ & $-0.03\pm0.18$ & $-0.02\pm0.10$ & $-0.02\pm0.15$ \\
25 & $0.04\pm0.19$ & $-0.19\pm0.33$ & $0.04\pm0.18$ & $-0.09\pm0.33$ & $-0.07\pm0.25$ & $-0.07\pm0.35$ & $-0.07\pm0.24$ & $-0.04\pm0.35$ \\
26 & $0.12\pm0.43$ & $-0.11\pm0.55$ & $0.12\pm0.42$ & $-0.11\pm0.63$ & $-0.17\pm0.51$ & $-0.09\pm0.59$ & $-0.17\pm0.51$ & $0.06\pm0.69$  \\
27 & $0.27\pm0.75$ & $0.22\pm0.85$  & $0.27\pm0.76$ & \nodata        & $-0.28\pm0.81$ & $0.40\pm0.90$  & $-0.25\pm0.82$ & \nodata        \\
\hline
\hline
\end{tabular}
\label{tab:sys}
\end{table*}

\subsection{Flags}\label{sec:flags}

In Table \ref{tab:data} we flag objects with suspicious (flag value
1), bad (2), or non-existent (3) fitting results.  All other objects
(good fits) have flag value of zero. Suspicious fitting results (with
flag value 1) are those where the \gf magnitude deviates from the
expected magnitude by more than three times the uncertainty in the \gf
magnitude derived as described below, in \S\ref{sec:random}.  This
comparison is illustrated in Figure \ref{fig:magmag}, where we show
the \gf and \sext magnitudes for the UDS F160W measurements.  The
expected magnitude is the BEST magnitude from \sext, corrected for the
systematic, magnitude-dependent offset between \gf model magnitude and
the BEST magnitude, indicated by the red line in the figure.  The BEST
magnitude is measured in $\mh$, adding the color, e.g., $\mj - \mh$,
measured over the $\mh$ segmentation map.  This offset between the
\sext and \gf magnitudes is 0.1 mag for bright sources and increases
to 0.3 mag for faint sources (see Figure \ref{fig:magmag}).  This
offset has been noted by many authors before \citep[e.g.,][]{holden05,
  blakeslee06, haussler07}.  The distribution of \gf fitting
parameters with flag value 1 is shown in red in Figure
\ref{fig:summary}.

Objects with flag value 1 are not necessarily bad fits and in some
cases can be used without problem.  However, we recommend to assess on
an object-by-object basis whether the results can be used and to
examine in those cases the \gf model and residual, which are also
released along with this paper (\S\ref{sec:sum}).

Bad fitting results (with flag value 2) are those for which one or
more parameter reached the constraint value forced onto \gf.  Note
that these fits may in fact truly represent the best possible \ser
profile, but that the inferred structural parameters are in most cases
not astrophysically meaningful.  This should be assessed on an
object-by-object basis and refined, hand-tuned fitting.  The user may
adopt inferred structural parameters at their own risk.  Objects that
are fit simultaneously with the target galaxy are allowed to reach
those constraint values -- the flag value for the target remains 0 in
this case.  A flag value 2 is also assigned in case the axis ratio
drops below 0.1, which \gf tends to do in the case of seemingly
converged, yet bad fits of generally faint objects.  The distribution
of \gf fitting parameters of objects with flag value is shown in blue
in Figure \ref{fig:summary}.  Finally, non-existent results (with flag
value 3) are simply those fits that `bombed', in \gf parlance.

The depth of our catalog of structural parameters for objects with
flag value zero is $\mh\sim 25$ as can be seen in the $\mh$ magnitude
histogram in Figure \ref{fig:summary}.  Accuracy and precision of the
measurements for flag-zero objects are discussed at length in
\S\ref{sec:err}, but this canonical depth of $\mh\sim 25$ is two
magnitudes shallower than the $5\sigma$ limit for CANDELS `wide'
\citep{koekemoer11}.  Thus, for objects in the range $25 < \mh < 27$
that are securely detected by CANDELS `wide' a measurement of even
their basic structural parameters will typically not be possible.
There is a population of flag-zero objects that extends to fainter
magnitudes, up to $\mh\sim 28$ (most easily seen in the
magnitude-radius panel of Figure \ref{fig:summary}); these objects are
located in the UDF.

\section{Measurement Uncertainties}\label{sec:err}

\subsection{Random Uncertainty Estimates from Internal
  Comparison}\label{sec:random}
A robust way to assign random uncertainties to the \gf measurements is
to compare measurements of the same objects in different data sets.
For the objects detected in the `deep' region of GOODS-S, we rerun
\gala on a shallower mosaic that has the same depth as the `wide'
CANDELS imaging.  All steps are performed in a manner that is
precisely analogous to that applied to other data sets.  First, the
appropriate hybrid PSF model is constructed as described in
\S\ref{sec:psf}, and \gala is used to estimate the background level,
as described in \ref{sec:bg}, before running \gf.  The input
segmentation map and \sext catalogs are identical for the for the
`deep' and the `shallow' versions of the mosaics.  Therefore,
neighboring objects are treated in exactly the same way in both cases,
and differences in \gf fitting results are entirely driven by the
noise and background properties of the images such that the variation
between the two sets of measurements reflect the uncertainty.  Note
that when running \gf we do not distinguish between pixels included
and excluded by the segmentation map, which is only used to identify
objects.

The measurement uncertainties in the structural parameters depend
foremost on the $S/N$, as can be seen in Figure \ref{fig:sn_par}.
However, there are several complicating factors that significantly
affect the true uncertainties: 1) the uncertainties in different
parameters are correlated (see Figure \ref{fig:correrr}); 2) after
normalizing for $S/N$, the uncertainties themselves also depend on the
parameter values in a non-trivial manner (see Figure
\ref{fig:correrr2} for a demonstration), and; 3) the covariance
between the measurement uncertainties changes with the values of the
parameters and their respective uncertainties.

We adopt an empirical approach in which the two fitting results (from
the deep and the shallow data) of sets of similar galaxies are used to
calculate, throughout parameter space, the uncertainties in each of
the measurement parameters.  We assume that the measurement
uncertainties depend on $m$, $n$, and $r$ (see Figure
\ref{fig:correrr2}), but not on any other parameter: we do not see
evidence for a correlation between $q$ (or, more obviously, $PA$) and
the uncertainties in any of the parameters.

We construct a parent sample of 6492 galaxies $\vec{g}$ with \gf
measurements from deep and shallow data and a flag value less than two
(see Sec \ref{sec:flags}) for both fits.  For each galaxy $g_i$ we
identify the 200 most similar galaxies in the parent sample $\vec{g}$
by computing the normalized distances $\vec{p_i}-\vec{p_j}$ in the
three-dimensional parameter space spanned by $m$, $n$, and $r$.  That
is, $\vec{p_i}$ is defined as
\begin{equation}
  \vec{p_i}=(m_i/\sigma(\vec{m}),~ \log{n_i}/\sigma(\log{\vec{n}}),~
  \log{r_i}/\sigma(\log{\vec{r}})),
\end{equation}
where $\sigma$ denotes the standard deviation in the respective
parameters. These factors are introduced in order to make differences
in parameter values comparable and produce dimensionless, normalized
distances.

For the 200 galaxies\footnote{This number can be changed by a factor
  of two without significantly changing the results.\label{fn}} most
similar to the target galaxy $g_i$, as defined by $\vec{p_i}$, we
compute the 16-84\%-tile range in the differences between the
parameters inferred from the deep and the shallow imaging.  This
provides an uncertainty estimate for each parameter as a function of
the parameters $m$, $n$, and $r$:
\begin{equation}
  \vec{\delta_i}=\vec{\delta}(\vec{p_i}) = (\delta m_i,~ \delta \log{n_i},~ \delta \log{r_i},~ \delta q_i,~ \delta PA_i)
\end{equation}

$\vec{\delta_i}$ is normalized through multiplying by $S/N_i$, as
measured from the segmentation map (see \S\ref{sec:gf}); this removes
the first-order dependence on $S/N$ and allows us to compute the error
budget for galaxies from images with different exposure
times\footnote{Here, we add the noise from the deep and shallow images
  in quadrature. Thus, rather than adopting the measurement from the
  deep image as `truth', we compute the combined error from the two
  measurements.}.  By repeating this computation for all galaxies
$\vec{g}$ we map out the measurement uncertainties throughout
parameter space sampled by $\vec{p}$.

This large matrix serves as a database which we use to estimate the
uncertainties in the \gf parameters for all galaxies in all our
fields.  As before, we calculate the distances $\vec{p_i}-\vec{p_j}$,
where $\vec{p_j}$ represents the objects in the database, and
$\vec{p_i}$ represents the galaxies to which we want to assign
measurement uncertainties.  We take the average $\vec{\delta_j}$ of
the 25 `nearest' galaxies $g_j$ in the database\footnote{see Footnote
  \ref{fn}}, and divide by $S/N_i$ to provide $\vec{\delta_i}$ with
the appropriate amplitude.  This quantity represents the measurement
uncertainties in all parameters ($m_i$, $n_i$, $r_i$, $q_i$, and
$PA_i$).  Note that the uncertainties are correlated with each other,
approximately as shown in Figure \ref{fig:correrr}.  This figure
serves as a mere illustration; not only the amplitude of the
uncertainty but also the covariance depends on the measurement
parameters themselves (i.e., on $\vec{p_i}$).

All measurements and their uncertainties (marginalized over all other
parameters are described above) are provided in Table \ref{tab:data}.
For uncertainties that are computed in logarithmic units we give
uncertainties in the customary linear units in the table, where
$\delta n = \ln(10)~n~\delta \log{n}$ in the case of, for example, the
\ser index.  The magnitude and its uncertainty are kept in the usual
archaic form. For reference, average random uncertainties (as well as
the systematic uncertainties -- see below, in \S\ref{sec:sys}) are
given in Table \ref{tab:sys} for galaxies with different magnitudes,
sizes, and \ser indices.  The bottom line is that for the CANDELS
`wide' imaging in F160W the parameters $m$, $r_e$ and $q$ can be
inferred with a random accuracy of 20\% or better for galaxies
brighter than $\mh\sim 24.5$, whereas $n$ can be measured at the same
level of accuracy for galaxies brighter than $\mh\sim 23.5$.

For those readers who wish to derive their own uncertainty estimates
(for example, in a difference confidence interval, or for
investigating asymmetric uncertainties) we provide our fitting results
for the `wide' sample in the GOODS-South, which consists of 6492
objects (Table \ref{tab:wide}).  The IDs of these objects match those
in the F160W catalog for GOODS-South.

While the uncertainty matrix is generated from F160W data, we can also
use it to assign uncertainty estimates for the structural parameters
measured in the other filters, again using the $S/N$ to provide the
uncertainties with the correct amplitude.  Several factors may
compromise this approach: the generally less smooth light distribution
at shorter wavelengths will tend to increase in the uncertainties
associated with single-component \ser fits, while the smaller PSF will
tend to decrease the uncertainty, especially for compact objects.
However, uncertainties depend to first order on $S/N$ alone and there
is no reason to assume that the correlations shown in Figures
\ref{fig:correrr} and \ref{fig:correrr2} substantially change with
filter choice.

We note that the reported uncertainties represent the convolved
contributions from noise and the uncertainty in the background level,
and we refer to \citet{guo09} and \citet{bruce12} for detailed
analyses of errors due to uncertainties in the background flux.  To
provide an indication of the contribution of the uncertainty in the
background we rerun \gf for 1000 randomly picked galaxies with
estimated background values $b_i$ in the UDS, assigning background
levels $b_j$ as calculated by \gala for another 1000 randomly picked
galaxies.  The standard deviation in $b_j-b_i$ is $1.4\times
10^{-4}~\rm{e}/\rm{s}/\rm{pix}$.  We then rerun \gf to obtain 1000
sets of parameter estimates that we compare with the original
estimates.  Because the variation in background is likely partially
real, the variation in the structural parameter estimates provides an
upper limit on the contribution of background uncertainties to the
measurement errors.  This exercise shows that at most $\sim 25-30\%$
of the total error budget as derived above is due to uncertainties in
the background flux estimates.  Hence, for most objects the accuracy
in the structural parameter estimates is limited by the $S/N$, and not
the background flux level, even though the latter is not entirely
negligible.  Only for very faint sources ($\mh>25.5$) with large sizes
($r>0.4$\arcsec) the uncertainty in the magnitude and structural
parameter estimates starts to be dominated by the uncertainty in the
background estimate, but for such faint sources the measurements are
highly uncertain anyway (see \S\ref{sec:sys}).

\begin{table}\scriptsize
  \caption{ 
    \gf fitting results for the F160W `wide' sample in GOODS-South (6492 objects) used for estimating the uncertainty estimates 
    through comparison with the `deep' fitting results (see \S\ref{sec:random}).  
    The ID matches the ID in the general catalog for GOODS-South.
    An electronic version of the full catalog is published online.  For further explanation of the table entries, see Table \ref{tab:data}.
  }
\begin{tabular}{ccccccc}
  \hline
  \hline
  ID & $m$ & $r$ & $n$ & $q$ & $PA$ & $S/N$ \\ 
  \hline
  & AB mag & $\arcsec$ & & & $\deg$ & \\
  \hline
2989 & 24.09 & 0.18 & 1.16 & 0.53 &  29.84 & 23.15 \\
3034 & 24.05 & 0.25 & 0.56 & 0.47 & -51.53 & 23.71 \\
3113 & 22.79 & 2.12 & 7.15 & 0.53 &  82.94 & 20.68 \\
3148 & 23.51 & 0.16 & 1.85 & 0.27 & -68.60 & 42.53 \\
3166 & 25.17 & 0.43 & 0.71 & 0.77 &  24.50 &  4.68 \\
   . & . & . & . & . & . & . \\
   . & . & . & . & . & . & . \\
   . & . & . & . & . & . & . \\
  \hline
  \hline
\end{tabular}
\label{tab:wide}
\end{table}

\subsection{Systematic Uncertainties from Simulated
  Images}\label{sec:sys}

From the data itself it is difficult to infer systematic uncertainties
as the `true' light distributions of the galaxies are unknown.  We use
simulated images with galaxies with known light distributions to
quantify these.  The simulated images are generated as described by
\citet{haussler07}, with the background noise taken directly from
empty parts of the UDS F160W mosaic.  The input \ser profiles of the
fake galaxies have the same magnitude, size and shape distributions as
in the real data.  The input profiles are convolved with the same PSF
as used in the data analysis -- this exercise is not designed to test
for systematic effects due to errors in our PSF model.  The simulated
mosaic is then processed as described in \S\ref{sec:data} and the
inferred structural parameters can be compared with the input values.

In Table \ref{tab:sys} we give the average difference between the
input and output values of the various parameters, along with the
sample average random uncertainties assigned as described above in
\S\ref{sec:random}.  For the largest part of parameter space, and
especially for the magnitude range of interest for morphological
studies of massive, high-redshift galaxies, the systematic differences
are substantially smaller than the random uncertainties; that is,
random uncertainties dominate.  However, systematics are not
negligible, especially for objects with large \ser~index.  Generally
speaking, structural parameters can be measured with a precision and
accuracy better than 15\% down to $\mh\sim 23$. At fainter magnitudes,
there are larger systematic and random uncertainties, especially for
large, high \ser index objects.  Since the typical faint,
high-redshift galaxies is small and has a low \ser index, 10\%-level
accuracy and precision in the basic size and shape parameters should
be reached down to $\mh \sim 24.5$.

Although the tabulated values strictly apply to the `wide' imaging,
the small changes per magnitude bin imply that the systematic effects
to the `deep' imaging are not substantially different.  We encourage
all users of the public catalogs to take the results given in Table
\ref{tab:sys} into account in their analysis.  We do not correct the
measurements for these systematic effects, as we cannot quantify the
precise amount for each individual galaxy.  The results presented here
serve as indications of the magnitude of the systematic uncertainties.
Note that these systematic uncertainties do not include errors in the
WFC3 zero points or other uncertainties in the calibration.

The variance in the difference between the input and output structural
parameters reflect random uncertainties that are very similar to those
inferred in \S\ref{sec:random}, with a similar dependence on
$S/N$ and correlation between the various parameters.  The reason that
we choose to use real rather than simulated data to infer our random
uncertainties is that single \ser  profiles do not (necessarily)
represent the true light distribution of a galaxy very well, which
could lead to underestimated uncertainties in the case that idealized
simulated galaxies are used in the analysis.

As mentioned above, the input structural parameters for the
simulations are based on the observed distribution of parameters.
This introduces the risk that regions of parameter space with large
systematic effects are unjustifiably ignored by design.  We test the
degree to which very small galaxies can be recovered by our method by
generating an additional set simulated of galaxy images.  The input
magnitudes are in the range $23.5 < \mh < 24.5$, corresponding to the
faintest galaxies that we still have fairly precise measurements for
(see Table \ref{tab:sys}). The structural parameters are chosen to lie
on a linearly-spaced grid in $(\log(r),~n,~q,~PA)$ space, with the
semi-major axis $r$ ranging from 0.3 to 30 pixels, probing a much
wider range than observed.

We find that 95\% of all simulated objects with semi-minor axis
lengths of 0.3 pixels are recovered correctly, meaning that \gf does
not crash or converges to an incorrect value, and that there are no
systematic differences between the input and recovered parameter
values.  We conclude that the lack of large numbers of such small
galaxies in the real catalogs (as seen in, e.g., Figure
\ref{fig:correrr2}) is not due to limitations in spatial resolution.

\section{Public Data Release}\label{sec:sum}
This paper is accompanied by the public
release\footnote{http://candels.ucolick.org/} of a
number of materials.  Most importantly, we present catalogs containing
109,533 unique objects in three fields (GOODS-South, UDS, and COSMOS),
with structural parameter measurements for each object in at least two
filters, F160W and F125W, and, in some cases a third (F098m or F105W;
see \S\ref{sec:data}).  Catalogs for two additional fields
(GOODS-North and EGS) will be released upon completion of the survey,
as indicated in Table \ref{tab:list}.

The catalogs contain an identification number, coordinates measured
from the F160W mosaics (i.e., the coordinates are the same in the
catalogs for the different filters), a flag indicating the
availability/reliability of the \gf model (\S\ref{sec:flags}), the
structural parameters measurements (\S\ref{sec:gf}) and their random
uncertainties (\S\ref{sec:random}).  A sample of the UDS F160W catalog
is given in Table \ref{tab:data}, and the parameter distribution is
shown in Figure \ref{fig:summary}.  Image cutouts, model surface
brightness profiles, and residuals from the model fit are also made
available online, an arbitrarily chosen sub-sample of which are shown
in Figure \ref{fig:stamps}.  F160W object segmentation maps (see
\S\ref{sec:det}) are made available for each field, and PSF models,
constructed as described in \S\ref{sec:psf}, are published for each
field and filter.  These models combine stacked stars and \TT models
to provide optimal sampling and fidelity.

The structural parameter estimates are not corrected for systematic
uncertainties, which we quantify in \S\ref{sec:sys} and Table
\ref{tab:sys}, but we encourage users to verify the impact of
systematic errors on their analysis.  Table \ref{tab:sys} also
contains average random uncertainties for a range of magnitudes.  This
table serves as a guideline when choosing a magnitude limit, or
verifying the accuracy and precision of the fitting results for a
particular set of objects.
 
\acknowledgements {This work is supported by {\it HST} grant GO-12060.
  Support for Program number GO-12060 was provided by NASA through a
  grant from the Space Telescope Science Institute, which is operated
  by the Association of Universities for Research in Astronomy,
  Incorporated, under NASA contract NAS5-26555.}

\bibliographystyle{apj}

\end{document}